\documentclass[aps,pra,twocolumn,longbibliography]{revtex4-1}
\usepackage{epsfig,amsopn}
\usepackage{graphicx}
\usepackage{color,xcolor}
\usepackage{amsmath,amssymb}
\usepackage{enumerate}
\newcommand\bea{\begin{eqnarray}}
\newcommand\eea{\end{eqnarray}}
\newcommand\beq{\begin{equation}}
\newcommand\eeq{\end{equation}}

\def\nn{\nonumber}
\def\f{\frac}

\def\ep{\epsilon}

\def\ga{\gamma}

\def\ra{\rangle}

\begin{document}

\title{Decoherence in electron transport: back-scattering, effect on interference and rectification } 
 \author{ Abhiram Soori}  
 \email{abhirams@uohyd.ac.in}
 \affiliation{School of Physics, University of Hyderabad, C. R. Rao Road, Gachibowli, Hyderabad-500046, India}
\author{ Udit Khanna} 
 \affiliation{Department of Physics, Bar-Ilan University, Ramat Gan 52900, Israel}
\begin{abstract}
  Decoherence is an undesirable, but ubiquitous phenomenon in quantum systems.   Here, we study the effect of partial decoherence, induced via a B\"uttiker probe, on two-terminal electronic transport across one-dimensional quantum wires and rings, in both the linear and non-linear regimes. We find that dephasing causes backscattering when introduced locally in a ballistic channel. Further, we find that decoherence results in rectification when inversion is broken in the two-terminal transport set-up by a combination of a local dephasing centre and a static impurity.   Interestingly, the rectification strength and even its direction varies strongly with the relative distance between the probe and the scatterer.   We further analyze how decoherence affects characteristic quantum effects in electronic transport, such as, Fabry-P\'erot oscillations in  double-barrier setups, and Aharonov-Bohm interference in one-dimensional rings, and find that the amplitude of oscillations in conductance   is reduced by decoherence. 
\end{abstract}
%\pacs{}
\maketitle

\section{Introduction}

Quantum effects appear in electronic transport through a number of remarkable phenomena,
such as Fabry-P\'erot oscillations and Aharonov-Bohm interference, that may be readily observed
in appropriately designed mesoscopic devices.
Beyond specific observations, perhaps the strongest manifestation of quantum dynamics is that transport
through such devices can often be described (at least qualitatively) through a unitary scattering matrix relating
the wavefunctions of incoming and outgoing modes~\cite{datta1995}.
This approach, pioneered by Landauer and B\"uttiker, relies on the electrons maintaining phase coherence
while traversing through the device~\cite{landauer1957r,buttiker1988}.
However, phase information is inherently lost to some extent in any such device due to inelastic scattering, 
and it is important to understand the effects of this decoherence on quantum transport.
Various quantitative effects of dephasing, such as the suppression of interference patterns, have been 
studied previously~\cite{ady1990,imry,Brouwer1997,Aharony2012}.
%More recently, it has been realized that under the right conditions dephasing may also lead to surprising qualitative changes such as rectification effects in two-terminal devices~\cite{segal2013,bredol}. 
While rectification in two-terminal setup can be achieved by inelastic scattering when accompanied by broken time reversal symmetry [8,9], we show that rectification does not need breaking of time reversal symmetry. Instead, dephasing (by B\"uttiker probe method) can result in rectification when parity is broken.

The unitarity of the scattering matrix implies that the current in a two-terminal device should  
reverse its sign when the bias direction is reversed, i.e., the charge transport must be reciprocal in 
the absence of dephasing. 
Within the regime of linear response, this restriction is a special case of the Onsager-Casimir relations, 
which reflect the reversibility of the microscopic scattering processes~\cite{casimirRMP,imry}, 
and continue to hold even in the presence of dephasing~\cite{buttiker1986m}. 
Beyond the linear regime, this symmetry constraint is {\it not necessarily respected} in the presence of inelastic 
scattering processes. 
It is interesting to understand the conditions under which such scattering may lead to a breakdown of 
reciprocity in nonlinear transport. 
Earlier works~\cite{butti1993,buttiker2004,boris2004,nonlinearABring1,nonlinearABring2} 
considered this question in different setups  that include a magnetic field, treating electron-electron interactions 
self-consistently and demonstrated that non-reciprocity 
may emerge if time-reversal symmetry is broken in interacting systems. 
In this work, we study a noninteracting model based on a quantum chain with local dephasing, and demonstrate that 
reciprocity may be violated even in the presence of time-reversal symmetry, by breaking inversion symmetry instead. 
Additionally, we  study the effect of (partial) dephasing on Fabry-P\'erot and Aharonov-Bohm interferences.

Microscopic modelling of the myriad interactions that may lead to phase-breaking is challenging even for 
small systems~\cite{moshe2020,Giamarchi2020,Giamarchi2022}, and various phenomenological approaches have been developed to study such effects. 
These include adding imaginary terms to the Hamiltonian, or to the state~\cite{gefen1984,efetov1995}, 
as well as by modelling wavefunction collapse events~\cite{bredol}. Decoherence can also be introduced by time varying stochastic on-site potential~\cite{leo2022}. 
In this work, we study the effects of decoherence on transport in the steady state through the B\"uttiker 
probe method~\cite{butti88}, wherein a fictitious third terminal is connected to the device. 
This terminal is assumed to be in equilibrium at a carefully chosen voltage, such that there is no net transfer of 
charge from the probe to the setup. 
The probe does accept electrons from the setup at energies above the probe voltage, and then reinjects them 
into the device at a lower energy and with an arbitrary phase. 
The energy difference is absorbed within the probe. 
As such, the probe models the effects of inelastic scattering as well as the loss of phase information. 
The idea of introducing dephasing locally through an external reservoir has been realized experimentally as well, in the context 
of a Mach-Zehnder interferometer~\cite{roul09}. 
From a different point of view, the probe may be considered as a measurement device that is continuously monitoring the 
average charge density at the site it is connected to. 
This is because the voltage on the probe is, in principle, an observable which depends on the local density distribution in the system~\cite{datta1995}. 

Here, we consider the effect of decoherence on two-terminal steady state transport through several setups. 
Throughout this work, all reservoirs are treated as (spinless) one-dimensional chains (see Fig.~1). 
We begin, in section II, with transport across a single site (connected to the probe), and demonstrate that the 
inelastic scattering introduced by the probe not only introduces backscattering in an otherwise pristine quantum wire, 
but also leads to two parallel channels of transport, one of which is coherent while the other is incoherent. 
In section III, we consider transport across a cavity, modelled as a linear tight-binding chain, featuring Fabry-P\'erot oscillations. 
We find that the probe acts as a selector, only affecting certain modes of the cavity and leaving the others intact. 
Then in section IV, we consider an Aharonov-Bohm ring, modelled as a 4-site tight-binding chain, and study how the dephasing 
introduced by the probe affects the interference pattern. 
Next, in section V, we consider a system with broken inversion symmetry, and show the emergence of rectification in presence of dephasing. 
Finally, we summarise and conclude in section VI.

\section{Backscattering}\label{sec:bs}

\begin{figure}[htb]
  \centering
 \includegraphics[width=7cm]{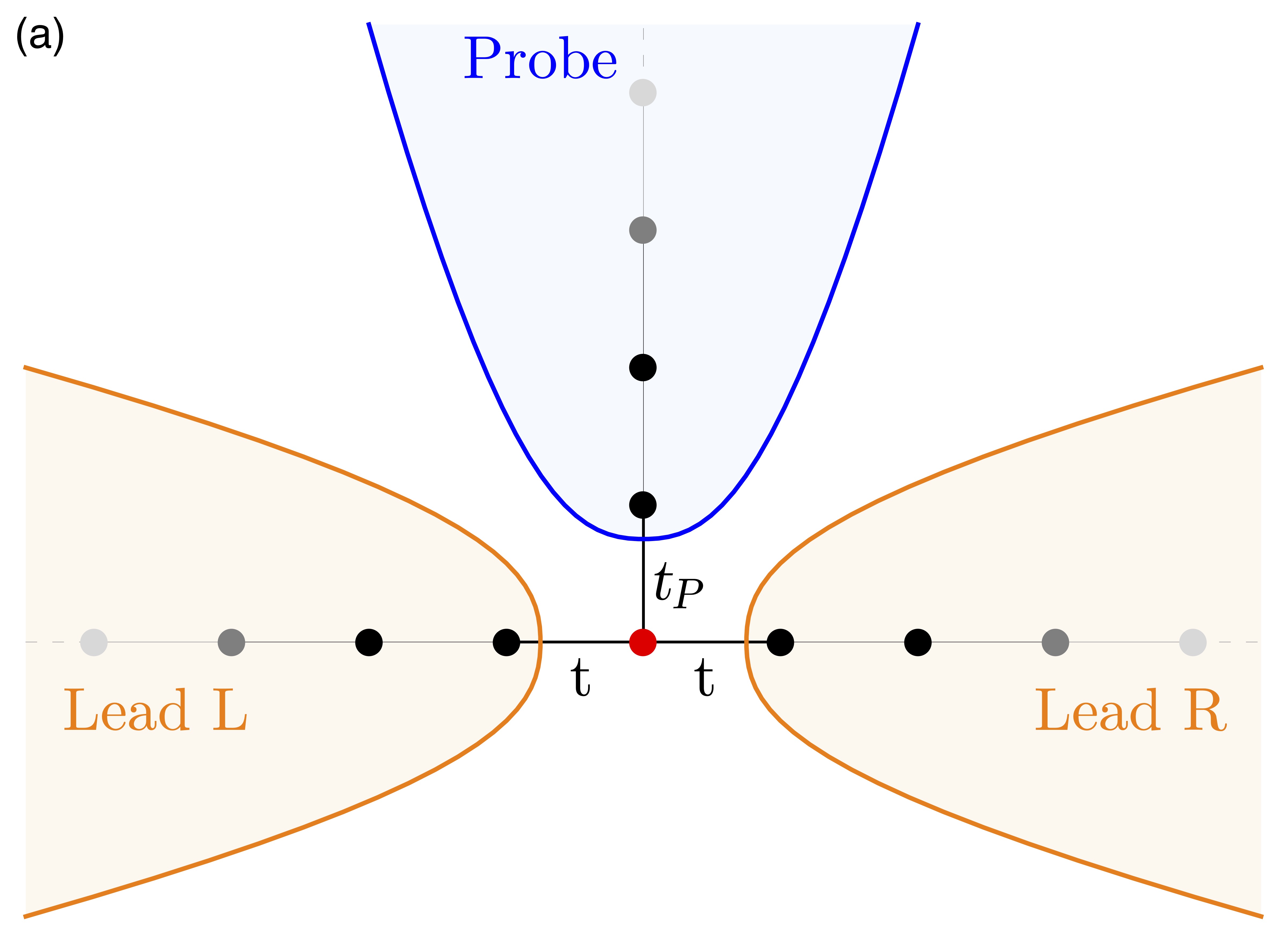}
 \includegraphics[width=\columnwidth]{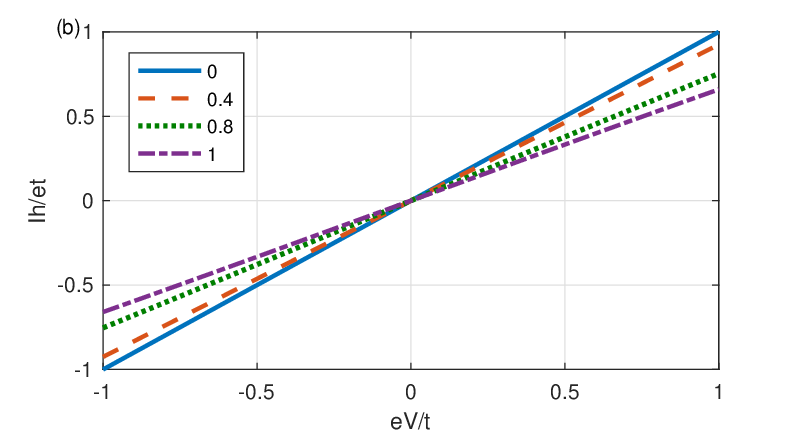}
  \caption{Current-voltage characteristics of a ballistic quantum wire in presence of decoherence. (a) The setup  
  comprises three reservoirs, modelled as one-dimensional chains with nearest neighbor hopping $t$, connected to a single site (marked in red). 
  Leads L and R are connected to the site with hopping amplitude $t$, forming a ballistic quantum wire, and act as the source and drain for the current. 
  While the voltage probe is connected with a hopping amplitude $t_{P}$ and introduces decoherence at the site, without drawing any net current from 
  the site. (b) I-V curves through the setup for different values 
  of $t_P/t$. The decreasing slope of the I-V curves with increasing $t_{P} / t$ is due to the backscattering introduced 
  by the probe.}\label{fig:backsca}
\end{figure}
A ballistic quantum wire connected to leads has a conductance of $e^2/h$ in absence of impurities, which also means that the transmission probability is unity~\cite{datta1995}. We introduce a B\"uttiker probe to the ballistic quantum wire connected to leads as shown in  Fig.~\ref{fig:backsca}(a) and study the current-voltage characteristics in Fig.~\ref{fig:backsca}(b). We find that the conductance is less than $e^2/h$ in presence of decoherence induced by the B\"uttiker probe, which implies that the decoherence induces backscattering in a ballistic channel.   

The Hamiltonian to study this phenomenon is
\bea 
H &=& -t\sum_{n=-\infty}^{\infty}(c^{\dag}_{n+1}c_n+~{\rm h.c.}) \nn \\ && 
-t\sum_{n=0}^{\infty}(d^{\dag}_{n+1}d_n+~{\rm h.c.}) -t_P(c^{\dag}_0d_0+{\rm h.c.}), 
\eea 
where $c_n$ annihilates an electron on site-$n$ in the ballistic quantum wire and $d_n$ annihilates an electron at site $n$ on the probe. Ballistic quantum wire extends from $n=-\infty$ to $n=\infty$, whereas the probe extends from $n=0$ to $n=\infty$. The hopping strengths in the wire and the probe are $t$. The bond that connects the wire to the probe has a hopping strength $t_P$.
We use the term ``source'' for the region $n\le -1$ on the quantum wire, and the term ``drain'' for the region $n\ge 1$ on the quantum wire. 
Dispersion in the wire and the probe are $E=-2t\cos{k}$. The scattering wavefunction for an electron incident on the wire from the left with energy $E$ has the form: $|\psi\ra = \sum_{n=-\infty}^{\infty}\psi_{n,w}|n,w\ra + \sum_{n=0}^{\infty}\psi_{n,p}|n,p\ra$ (the subscript $w/p$ refers to wire/probe), where 
\bea 
\psi_{n,w} &=& e^{ikn}+r_ke^{-ikn}, ~~{\rm for~~}n\le -1, \nn \\ 
&=& t_ke^{ikn},~~{\rm for~~}n\ge 1, \nn \\ 
\psi_{n,p}&=& t_{k,p}e^{ikn},~~{\rm for ~~}n\ge 0 , 
\eea
where $k=\cos^{-1}(-E/2t)$. Under an applied bias $V$ from source to drain, the currents in the source ($I_{SS}$), the probe ($I_{SP}$) and the drain ($I_{SD}$) are given by 
\bea  I_{SS}&=&e\int_0^{eV}dE(1-|r_{k}(E)|^2)/h, \nn \\ I_{SP}&=&e\int_0^{eV}dE|t_{k,p}(E)|^2/h, \nn \\  I_{SD}&=&e\int_0^{eV}dE|t_{k}(E)|^2/h.  \eea  
Now, a bias $V_P$ is applied at the probe so that the current flows from the probe into source and drain in such way that the net current in the probe is zero. The currents due to the bias $V_P$ on the probe into the source, drain and probe are  $I_{PS}$, $I_{PD}$ and $-I_{SP}$ respectively. These currents can be calculated in the following way. The wavefunction for an electron incident from the probe at energy $E$ can be written as 
$|\psi\ra = \sum_{n=-\infty}^{\infty}\psi_{n,w}|n,w\ra + \sum_{n=0}^{\infty}\psi_{n,p}|n,p\ra$,  where 
\bea 
\psi_{n,p} &=& e^{-ikn}+r_{k,p}e^{ikn} ~~{\rm for}~~n\ge 0  \nn \\
 \psi_{n,w} &=& t_{k,s}e^{-ikn} ~~{\rm for~~}n\le -1 \nn \\ 
 &=& t_{k,d}e^{ikn} ~~{\rm for~~}n\ge 1
\eea
Due to a voltage $V_P$ applied in the probe terminal, the current in the probe ($I_{PP}$), the current in the source ($I_{PS}$), the current in the drain $I_{PD}$ are given by: 
\bea
I_{PP} &=&  e\int_0^{eV_P}dE(|r_{k,p}(E)|^2-1)/h, \nn \\ 
I_{PS}&=&-e\int_0^{eV_P}dE|t_{k,s}(E)|^2/h \nn \\ 
I_{PD}&=&e\int_0^{eV_P}dE|t_{k,d}(E)|^2/h  
\eea
The voltage of the probe $V_P$ should be so chosen that $I_{PP}=-I_{SP}$. This choice of the probe voltage results in ``voltage probe''~\cite{segal2013}.  The total current in the quantum wire under the influence of the two biases is $I_{SS}+I_{PS}$. The net current in the probe is zero. The phases of the backscattered currents in $I_{SS}$ and $I_{PS}$ will be different resulting in dephasing. In the limit $t_P\ll t$, $I_{SP}$ becomes extremely small in magnitude, and the voltage $V_P$ chosen so that the total current in the probe terminal is zero, can be regarded as the voltage measured at site $n=0$ in the quantum wire. 
In Fig.~\ref{fig:backsca}(b), the numerically calculated current-voltage characteristics is plotted for different values of $t_P$. The I-V curve is found to be a straight line and  the conductance decreases with increasing $t_P$. 

 \begin{figure}[htb]
 \includegraphics[width=8cm]{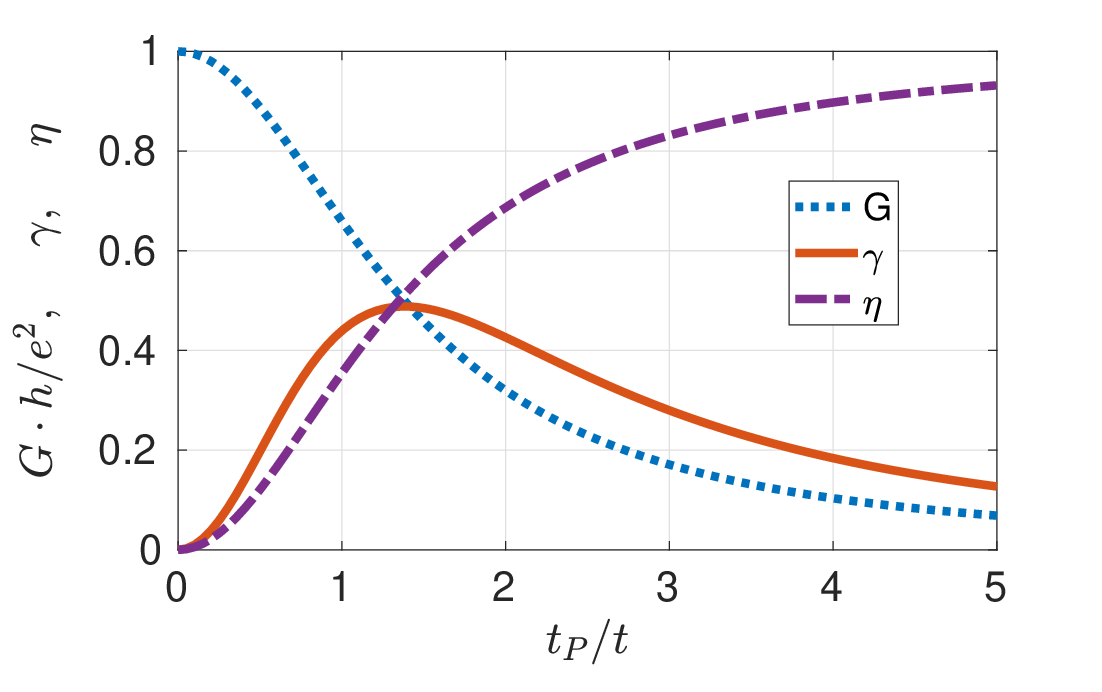}
   \caption{Variation of the conductance ($G$), dephasing factor ($\ga$) and coherent backscattering factor ($\eta$) with the 
   strength of connection to the probe ($t_P$). Increasing $t_{P}$ leads to higher backscattering and consequently to 
   a monotonically decreasing $G$. When the coupling to the probe is weak, the dephasing (quantified by $\ga$) increases with 
   $t_{P}$. On the other hand, for large values of $t_{P} / t$ a bound state localized at the junction appears in the spectrum, 
   which leads to higher coherent backscattering but smaller dephasing. 
   Consequently, the dephasing is nonmonotonic and maximal close to $t_{P} \approx t$. 
   These results are for bias $V=t/e$.}\label{fig:dephas}
\end{figure}

\subsection*{Extent of dephasing}
When a Buttiker probe is connected to a channel, a certain fraction of the incident current goes into the probe and an equal amount of current is supplied by the probe back into the channel. This fraction of current which is injected back into the the channel is responsible for dephasing. We define the ratio of the current injected back from the probe to the current incident from the source as dephasing factor: 
\beq \gamma=\f{(I_{PD}-I_{PS})h}{e^2V} \eeq  
A part of the current injected back into the channel from the probe travels back into the source, which is purely incoherent. Also, there is a coherent backscattering that happens when  electrons incident from the source scatter at the junction with the probe. We define coherent backscattering factor $\eta$ as the ratio of the coherently backscattered current to the total backscattered current, which is given by the expression: 
\beq \eta = \f{(e^2V/h-I_{SS})}{(e^2V/h-I_{SS}-I_{PS})} \eeq The conductance, $\gamma$ and $\eta$ are plotted versus coupling to the probe $t_P$ in Fig.~\ref{fig:dephas}. For large values of $t_P$, the two sites $n=0$ on the wire and $n=0$ on the probe form a dimer bound state and cause coherent backscattering. Thus, for large $t_P$, coherent backscattering is enhanced and dephasing is suppressed. Hence, the dephasing factor increases as $t_P$ is increased from $0$, reaches a maximum and then decreases. The conductance of the channel decreases monotonically with $t_P$, since the overall backscattering increases monotonically with $t_P$. At small $t_P$, both the coherent and the incoherent backscatterings contribute significantly, while at large $t_P$, the coherent backscattering dominates.

\begin{figure}[t]
  \centering
 \includegraphics[scale=0.2]{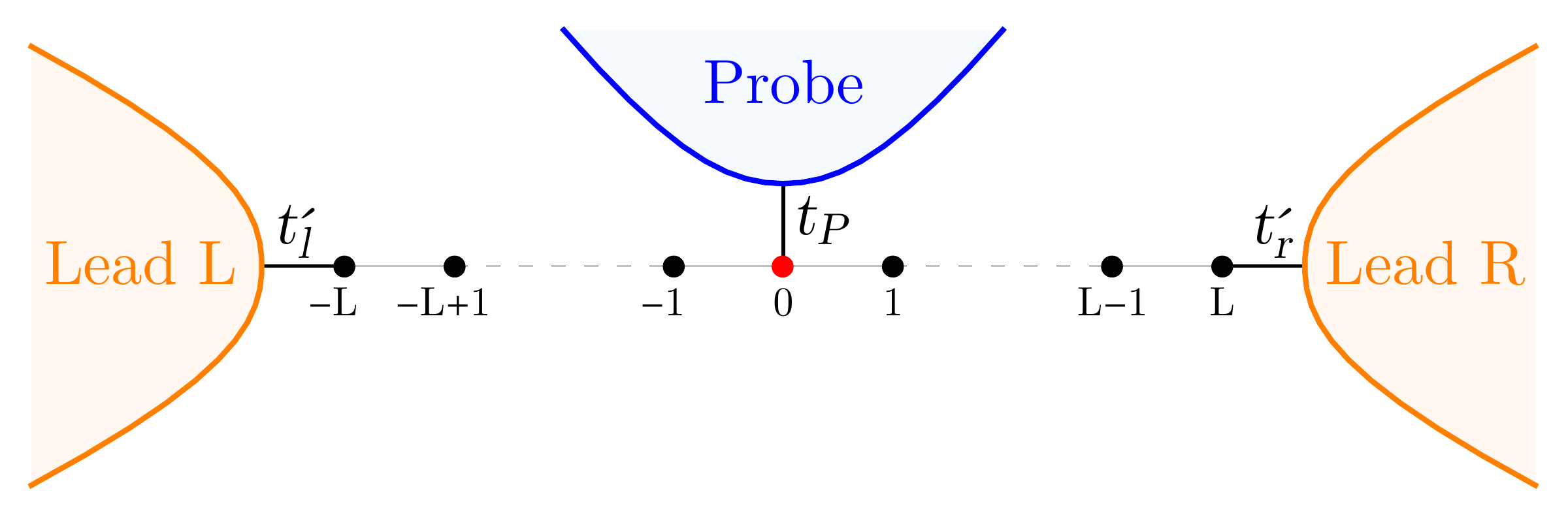}
  \caption{Schematic of a Fabry-P\'erot interferometer with decohering probe. The leads and the probe are modelled as 1D quantum wires 
  (as depicted in Fig.~\ref{fig:backsca}), and connected to a 1D cavity of length $2L+1$ with hopping amplitudes $t_{l}^{\prime}, t_{r}^{\prime}$ and $t_{P}$, 
  respectively. The probe is connected only to the central site of the cavity (marked in red) and hence, couples more strongly to the cavity 
  modes with higher weight at this site. It is assumed that the chemical potential $\mu$ inside the cavity can be controlled through an 
  external gate (not shown), which allows bringing different cavity modes within the bias window.}\label{fig:fabperosetup}
\end{figure}

\begin{figure}[htb]
 \includegraphics[width=8cm]{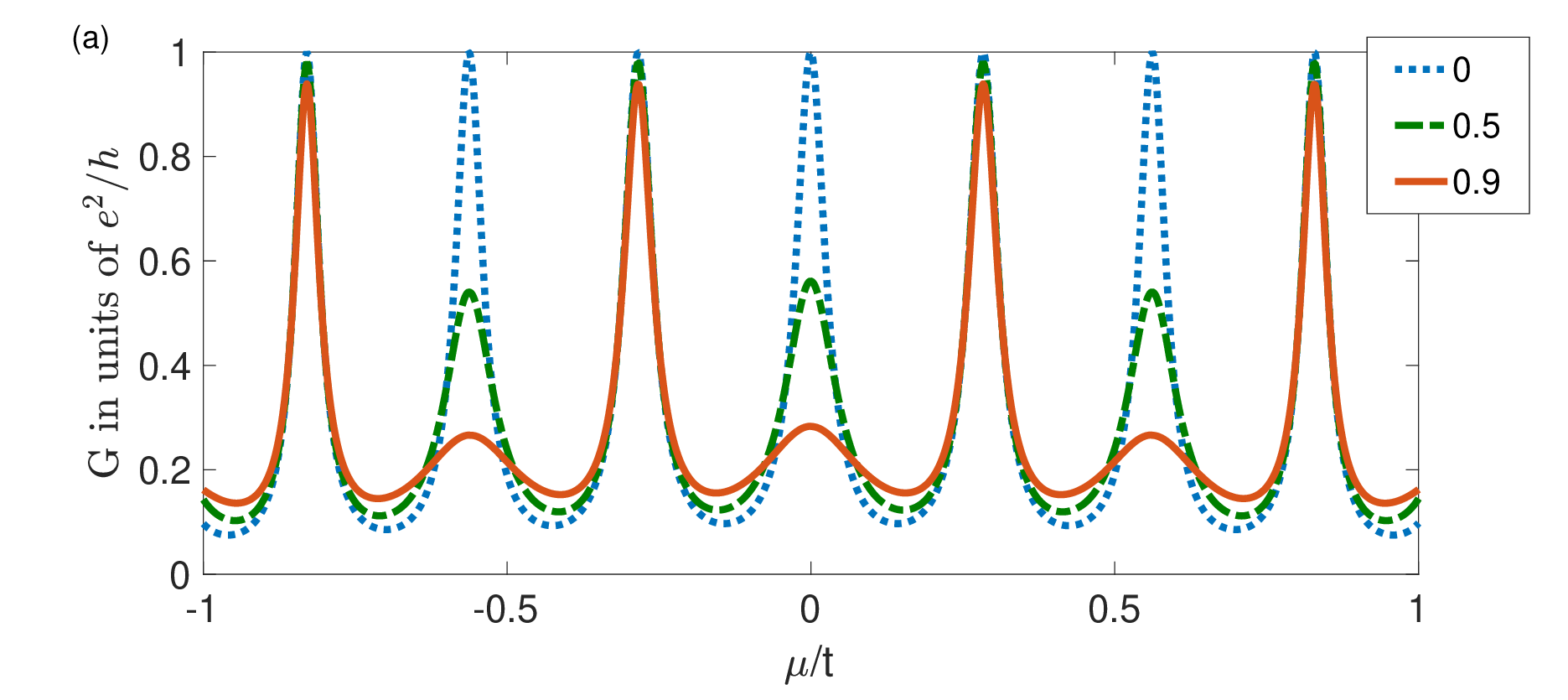}
 \includegraphics[width=8cm]{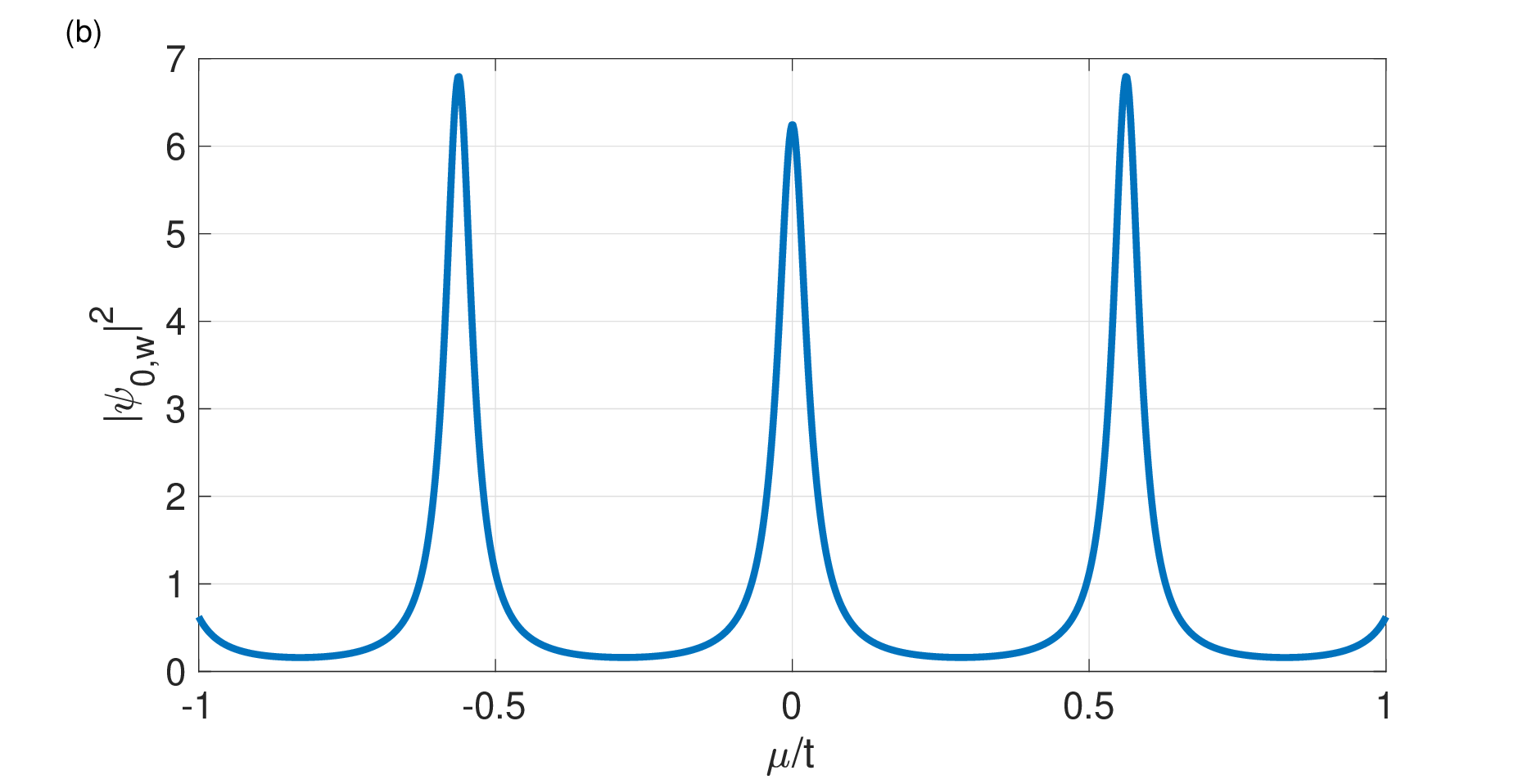}
  \caption{(a) $G=I/V$ for a small bias $V$ versus $\mu$. Different curves are for different values of $t_P/t$ shown in the legend. (b) $|\psi_0|^2$ for plane wave scattering in two-terminal device without the probe. The amplitude of oscillations decreases due to dephasing whenever $|\psi_0|^2$ is large. Parameters: $t'_l=t'_r=t'$, $t'/t=0.4$, $V=0.001t/e$, $L=10$. }~\label{fig:fabpero}
\end{figure}

\section{Effect on Fabry-P\'erot interference}
Let us consider a one dimensional channel of finite length weakly connected to leads on either sides.  If the chemical potential in the channel can be tuned (which can be achieved by the application of a gate voltage), the conductance of the channel for a small bias should oscillate with  a large amplitude when the chemical potential is changed~\cite{soori12,soori17,soori19,soori2021,soori22car}. This is the essence of Fabry-P\'erot type interference. Now, if dephasing is introduced in the channel at the centre, the amplitude of the Fabry-P\'erot oscillations decreases. This can be seen in Fig.~\ref{fig:fabpero}(a).  The Hamiltonian for the study of this setup is, % in eq.~\eqref{eq:ham-fp}. 
\bea  
H &=& -t\sum_{n=-\infty}^{-L-1} (c^{\dag}_{n-1}c_n + h.c.) -t_l^{\prime}(c^{\dag}_{-L-1}c_{-L}+h.c.) \nn \\ 
&& -t\sum_{n=-L}^{L-1} (c^{\dag}_{n+1}c_n + h.c.) -\mu\sum_{n=-L}^{L}c^{\dag}_nc_n \nn \\ 
&&-t_r^{\prime}(c^{\dag}_{L+1}c_{L}+h.c.)-t\sum_{n=L+1}^{\infty} (c^{\dag}_{n+1}c_n + h.c.)\nn \\ 
&& -t_P(c^{\dag}_0d_{1}+h.c.)-t\sum_{n=1}^{\infty}(d^{\dag}_{n+1}d_{n}+h.c.) \label{eq:ham-fp}
\eea
Here, $\mu$ is the chemical potential in the channel, which may be tuned through an external gate. 
The channel is connected to the source and drain leads through hopping matrix elements $t_{l}^{\prime}$ and 
$t_{r}^{\prime}$ which are assumed to be much smaller than the hopping $t$ within the chain. 
This models a sharp tunnel barrier at either end of the cavity. 
The method of calculating the conductance when dephasing is present is same as that in the previous section. 

When $t_P=0$, the scattering wavefunction for an electron incident from left with zero energy has the form
\bea 
\psi_{n,w} &=& e^{ikn}+r_ke^{-ikn}, {~~\rm for~~}n<-L, \nn \\ 
&=& s_+e^{ik'n}+s_-e^{-ik'n}, {~~\rm for~~} -L\le n\le L, \nn \\
&=& t_ke^{ikn}, {~~\rm for~~} n>L,
\eea
where $k=\pi/2$, $k'=\cos^{-1}[-\mu/2t]$. In Fig.~\ref{fig:fabpero}(b), $|\psi_{0,w}|^2$ is plotted versus $\mu$. 
At finite $t_{P}$, the amplitude of Fabry-P\'erot oscillations is suppressed largely at the values of $\mu$ for which $|\psi_{0,w}|^2$ is peaked (see, Fig.~\ref{fig:fabpero}(a)). This is because, dephasing affects the Fabry-P\'erot interference to a larger extent when the probability weight at the site of the dephasing is larger. 

\begin{figure}[htb]
  \centering
 \includegraphics[scale=0.2]{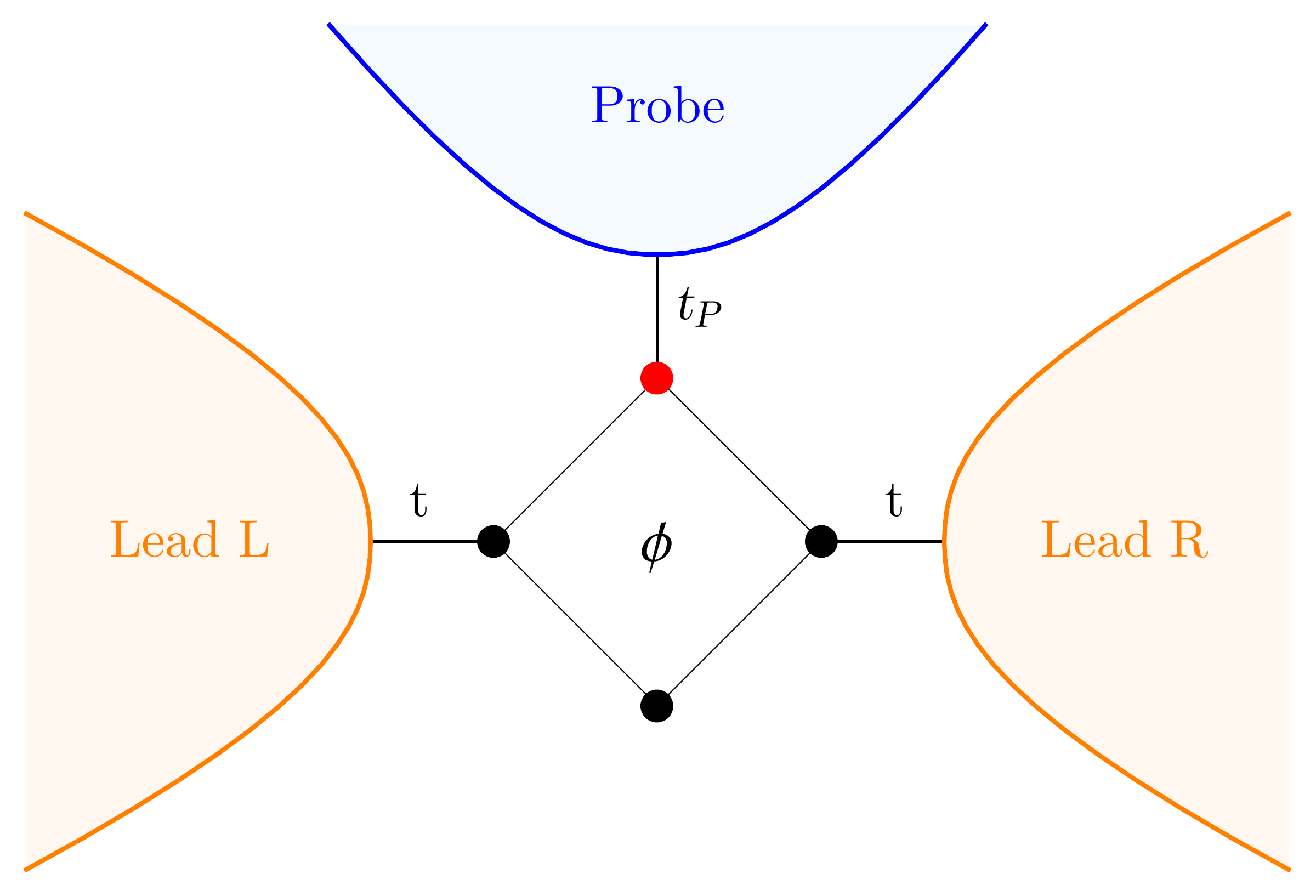}
  \caption{Schematic of an Aharonov-Bohm interferometer with a decohering probe. 
  The interferometer is modelled as a 4-site ring (with hopping amplitude $t$)  
  threaded by flux $\phi$. The source and drain leads are connected to two opposite sites of the ring with hopping amplitude $t$. 
  The probe is connected to a third site (marked in red) with amplitude $t_{P}$. All reservoirs are modelled as 1D chains
  (as depicted in Fig.~\ref{fig:backsca}).}\label{fig:ABsetup}
\end{figure}

\begin{figure}[htb]
 \includegraphics[width=8cm]{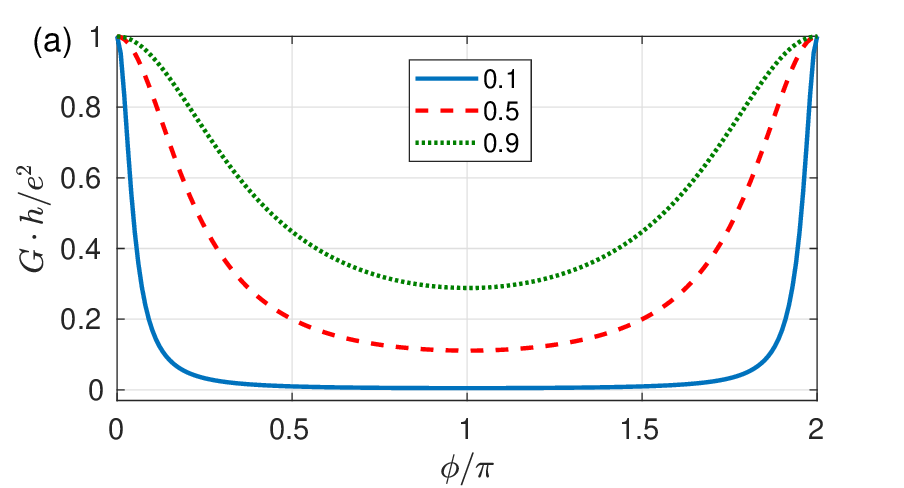}
 \includegraphics[width=8cm]{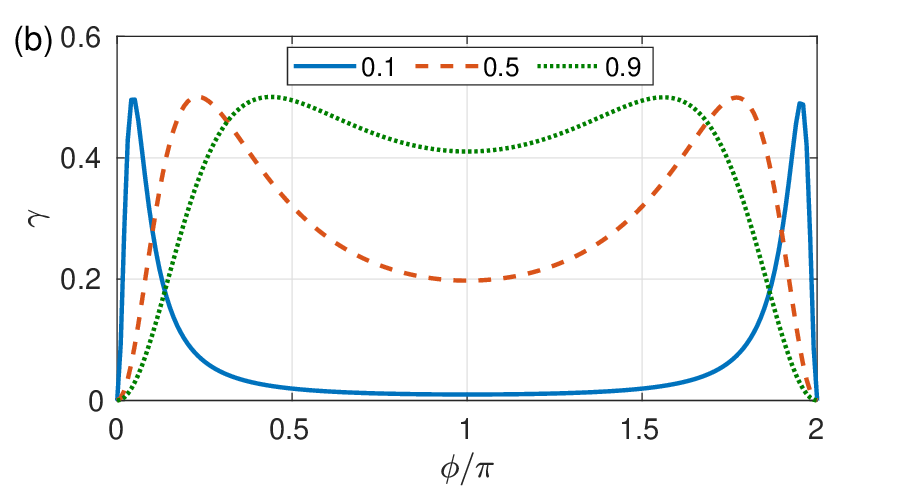}
 \caption{(a)Conductance versus the flux through the ring $\phi$ for different values of $t_P/t$ indicated in the legend. (b) Dephasing factor $\gamma$ versus $\phi$ for different values of $t_P/t$ indicated in the legend. Dephasing suppresses the amplitude of Aharonov-Bohm oscillations. }\label{fig:AB}
\end{figure}

\section{Effect on Aharonov-Bohm oscillations}
To study the effect of dephasing on Aharonov-Bohm oscillations, we connect a four-site ring threaded by a flux to leads on either sides. One of the sites on the ring is connected to a Buttiker probe as shown in Fig.~\ref{fig:ABsetup}. The strength of hopping onto the probe determines the extent of dephasing. The Hamiltonian to study the effect of dephasing on Aharonov-Bohm oscillations is 
\bea 
H &=& -t\sum_{n=-\infty}^{\infty} (c^{\dag}_{n+1}c_n + {\rm h.c.}) \nn \\ && -t(d^{\dag}_{0}c_{-1}e^{i\phi}+d^{\dag}_{0}c_1 + {\rm h.c.}) -t_P(d^{\dag}_{1}d_{0}+{\rm h.c.}) \nn \\ 
&&  - t\sum_{n=1}^{\infty} (d^{\dag}_{n+1}d_{n}+{\rm h.c.})
\eea

The current $I$ in response to a bias $V$ in this setup can be calculated using the same method as described in sec.~\ref{sec:bs}. The conductance $G=I/V$ for a small bias of $V=0.001t/e$ is calculated as a function of the flux $\phi$ and plotted in Fig.~\ref{fig:AB}(a).  In absence of dephasing, the current across the ring is zero for $\phi=\pi$ due to destructive interference. On the other hand the current is large for $\phi=0$.  Interestingly, at $\phi=\pi$, the current which is zero in purely coherent transport becomes nonzero as the dephasing ($t_P$) increases. This is in contrast to the case of ballistic channel, wherein dephasing suppresses conductance.  The conductance of the ring oscillates as a function of the flux $\phi$ in the range $(0,1)e^2/h$ for small values of $t_P$ for which the dephasing is small. The amplitude of such oscillations decreases with increasing $t_P$ for which the dephasing is larger. In Fig.~\ref{fig:AB}(b), the extent of dephasing $\ga$ is plotted as a function of $\phi$. The extent of dephasing is proportional to the current that flows through the probe when a bias is applied from source to drain. Such a current is zero for $\phi=0$ due to interference effect and increases as $\phi$ increases. Therefore, the dephasing factor increases as $\phi$ increases from zero. But as $\phi$ increases beyond a certain value in the range $0<\phi<\pi$, the current that flows from source to drain decreases. This makes the fraction of incident current that goes into the B\"uttiker probe decrease as $\phi$ increases beyond a certain value in the range $(0,\pi)$.  This results in a peak in the graph of dephasing factor versus $\phi$  in the range  $0<\phi<\pi$.

\begin{figure}[htb]
  \centering
 \includegraphics[scale=0.2]{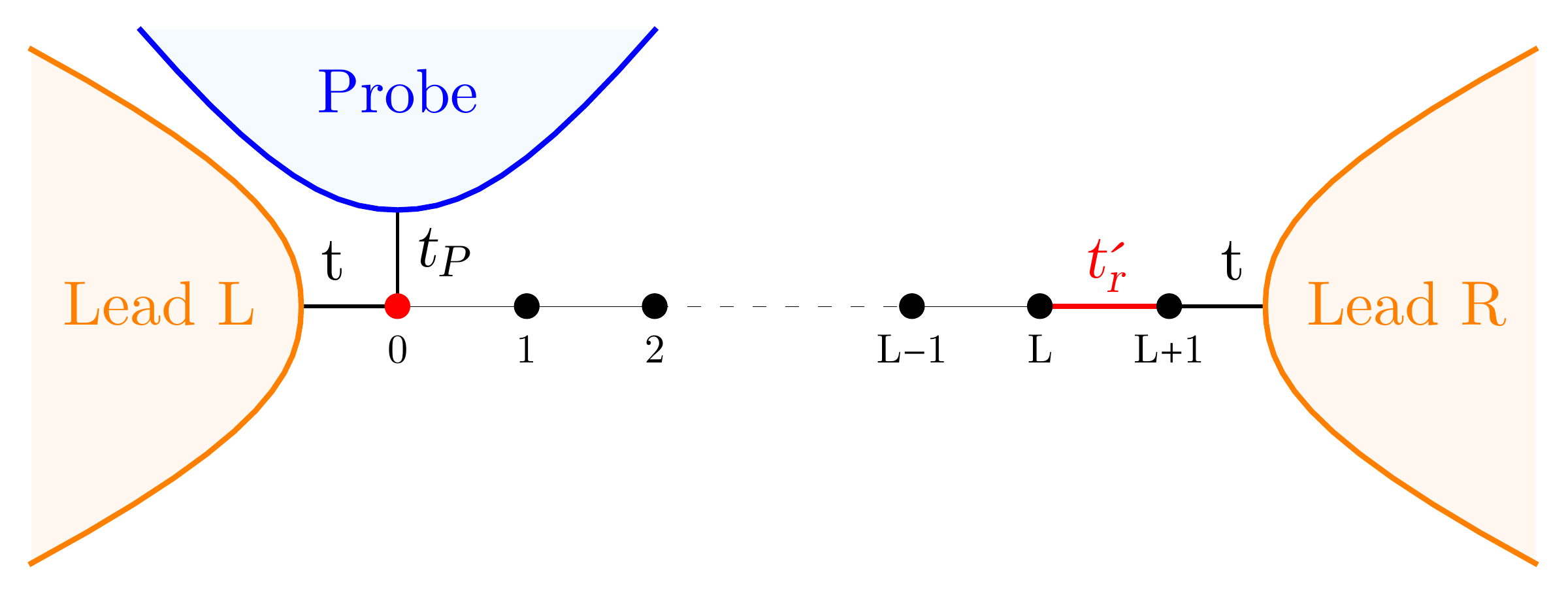}
  \caption{Schematic of a quantum wire with a scatterer, which features rectification in presence of decoherence induced by 
  a voltage probe. The scatterer is modelled as a disordered bond with hopping amplitude $t_{r}^{\prime}$. The probe is connected to a site 
  (marked in red) at a distance $L$ from the scatterer with hopping amplitude $t_{P}$. All other hoppings are equal and denoted by $t$. 
  The placement of the probe and the scatterer breaks the inversion symmetry of the setup and leads to finite rectification. 
  All reservoirs are modelled as 1D chains (as depicted in Fig.~\ref{fig:backsca}).}\label{fig:rectisetup}
\end{figure}

\begin{figure*}[htb]
 \includegraphics[width=5.5cm]{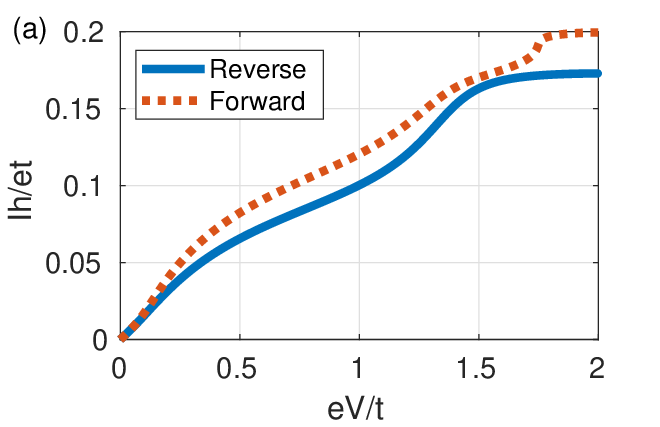}
 \includegraphics[width=5.5cm]{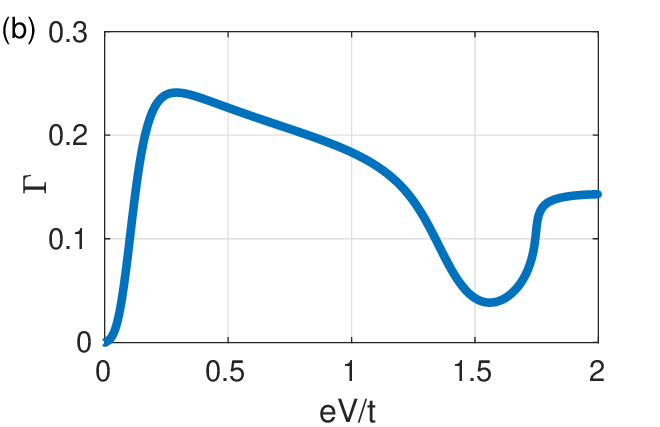}
 \includegraphics[width=5.5cm]{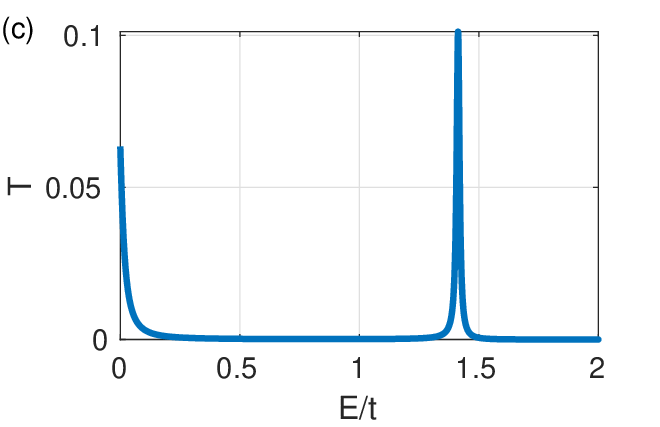}
 \includegraphics[width=5.5cm]{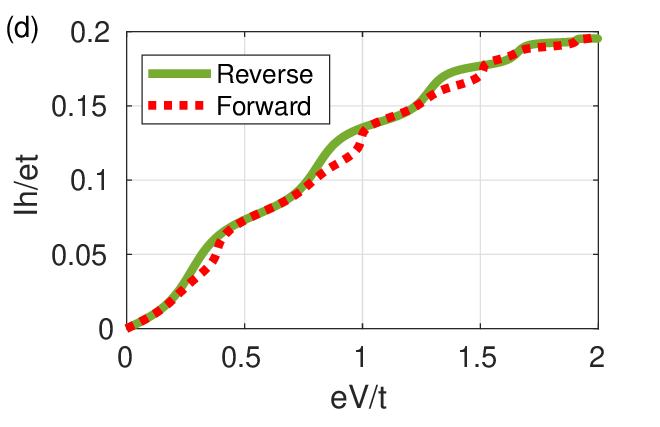}
 \includegraphics[width=5.5cm]{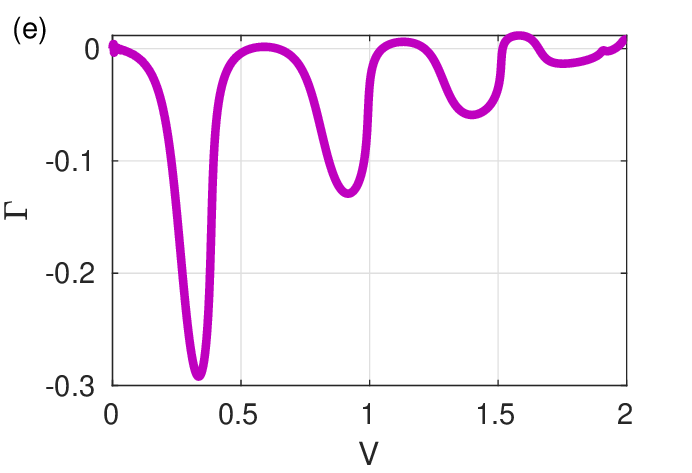}
 \includegraphics[width=5.5cm]{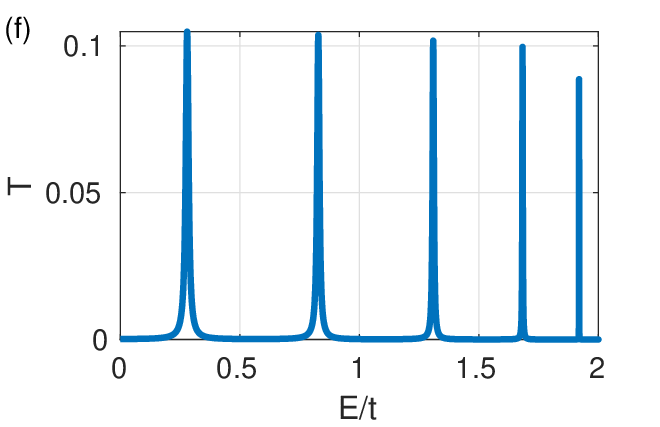}
  \caption{Rectification induced by dephasing.
  (a,d) I-V characteristics of the setup with current flowing in different directions. 
  Clearly, the current depends on the direction of the applied bias, indicating a rectification effect in the non-linear regime. 
  (b,e) Diode effect coefficient $\Gamma=2[I_F(V)-I_B(V)]/[I_F(V)+I_B(V)]$ as a function of the bias $V$, for the forward ($I_{F}$) and backward ($I_{B}$) currents depicted in (a,d) respectively. The rectification effect is not monotonic as a function of the bias, due to the interplay of the dephasing with the interference going on inside the cavity. 
  (c,f) Transmission probability of a coherent (but otherwise equivalent) cavity versus energy. 
  The local minima in $\Gamma(V)$ observed in (b,e) are seemingly correlated with the resonances observed in (c,f). 
  The length $L$ was assumed to be $3$ in (a,b,c) and $10$ (d,e,f). The other parameters are identical for all figures, 
  and taken to be $t'_r=0.2t$, $t_P=t$, $\epsilon_0=30t$.}\label{fig:recti}
\end{figure*}

\section{Diode effect}

Now, we study the possibility of diode effect in presence of dephasing. Breaking of parity is necessary for diode effect. We break parity by connecting site $0$ to a B\"uttiker probe and making the hopping on the bond between sites  $L$ and $(L+1)$ weak. We take the Hamiltonian in eq.~\eqref{eq:ham-fp} and choose $t'_l=t$, $\mu=0$. This setup is depicted in Fig.~\ref{fig:rectisetup}. The method of calculation of $I-V$ characteristics is same as the one explained in Sec.~\ref{sec:bs}. In Fig.~\ref{fig:recti}, we plot $I-V$ characteristics for forward and reverse biases, diode effect coefficient $\Gamma$ versus bias and the dephasing factor $\gamma$ versus bias for $L=3$ and $L=10$. Here, we choose $t'_r=0.2t$ and $t_P=t$. The hopping onto the probe $t_P$ is chosen to be $t$ so that the dephasing is large, which is expected to result in a higher rectification. In the limit $t_P\ll t$, the diode effect coefficient becomes exteremely small and there is no rectification. In principle, the choice $L=0$ also breaks parity. But the I-V characteristics for this choice of $L$ is linear and the diode effect does not show up.  The diode effect coefficient $\Gamma$ is defined by $\Gamma=2[I_F(V)-I_B(V)]/[I_F(V)+I_B(V)]$. 
The diode effect coefficient shows oscillations as the bias is varied. 
$k(L+1)=n\pi$, where $n$ is an integer,  is the Fabry-P\'erot interference condition. The values of the bias $V$ at which this condition is satisfied is $V=-2t\cos{[n\pi/(L+1)]}$. For $L=10$, these values of $V$ are  $0.28t$, $0.83t$, $1.3t$, $1.7t$, $1.9t$ for $n=6, 7, 8, 9, 10$ respectively. At these values of the bias, the diode effect coefficient is large in magnitude as can be seen from Fig.~\ref{fig:recti}(e).

To explain the features in Fig.~\ref{fig:recti}(b,e), we consider a two terminal setup where, on an otherwise ballistic wire with hopping $t$, the bond between sites $L$ and $L+1$ is made weak and also on site $0$, an onsite energy $\ep_0$ is introduced. There is no dephasing in this setup. Transmission probability for such a setup as a function of energy is shown in Fig.~\ref{fig:recti}~(c) and Fig.~\ref{fig:recti}~(f) for $t'_r=0.2t$, $\epsilon_0=30t$. Current as a response to a bias is proportional to the integral of transmission probability over the bias window. Dephasing by voltage probe results in inelastic scattering where the electron that crosses the dephasing centre loses energy. So, for $L=3$, when the bias $V$ is small, the electrons at energy $E>0$ incident from left to right first interact with the dephasing centre, lose energy and for the lower energy electrons the transmission probability is higher (as compared to the case without dephasing) as can be seen from Fig.~\ref{fig:recti}~(c). However, for the same bias window in the opposite direction, the electrons are incident from right to left. These electrons first interact with the resonant cavity, transmit to the left with a probability less than that for the forward bias and after the transmission, they interact with the dephasing centre. The electrons lose energy on dephasing and a fraction of the part that has undergone dephasing  gets again backscattered to the right. So, on the whole, for a small bias, the current transmitted from left to right is higher than that from right to left resulting in positive diode effect coefficient that increases with bias. But as the bias is close to $1.4t/e$, the diode effect coefficient decreases and shows a valley. This is because, as the bias approaches a resonant energy from left, the electrons incident from left to right lose energy before transmission through the cavity resulting in lower transmitted current compared to the electrons incident from right to left which get transmitted through the cavity first (with a higher probability) and then interact with the dephasing centre, losing energy and getting trapped on the left side. A reasoning on similar lines explains the features of Fig.~\ref{fig:recti}~(e).

\section{Discussion}
Suppression of amplitude of Fabry-P\'erot oscillations due to dephasing studied in this work  can be tested experimentally  by coupling a resonant cavity to a probe and applying a gate voltage to the cavity.  We could not find any work in the literature which studies the effect of dephasing on Fabry-P\'erot oscillations by tuning the chemical potential.  The effect of dephasing on Aharonov-Bohm interference is well-known~\cite{pareek1998,benjamin2002}. Our calculations qualitatively agree with these results, though the method followed differs in details.  While diode effect in non-interacting four-terminal transport is not uncommon~\cite{song98}, two terminal quantum transport does not result in diode effect. We find that introducing dephasing in the transport channel results in diode effect in two terminal set-up when inversion symmetry is broken. This is in contrast to other works where time reversal symmetry breaking is needed to get diode effect in presence of dephasing~\cite{bandyo2021,bredol,bedkihal2013prb}. Our finding that diode effect manifests in non-interacting systems in presence of dephasing when inversion is broken without the need of time-reversal breaking is a new finding. The idea of B\"uttiker probe to introduce dephasing in a channel in a controlled way has been experimentally realized~\cite{roul09}. In addition to a B\"uttiker probe, a barrier can be designed in experiments by an applied gate voltage to a small segment of a one-dimensional transport channel. Hence the set-ups proposed by us can be tested experimentally. 

\section{Summary and Conclusion}

In this work, we explore the impact of dephasing on electron transport in noninteracting systems using a B\"uttiker probe to introduce local dephasing. We observe that dephasing in a ballistic channel leads to backscattering.  At low coupling to the B\"uttiker probe, both coherent and incoherent contributions to backscattering are present, while at high coupling strengths, coherent backscattering becomes dominant. The dephasing factor $\gamma$ peaks at an intermediate coupling strength, whereas the coherent backscattering factor $\eta$ increases monotonically with coupling.

In a cavity, local dephasing results in the suppression of Fabry-P\'erot oscillations in conductance as the cavity's chemical potential varies. This suppression is more pronounced when the wave function's probability weight is significant at the dephasing center. When the B\"uttiker probe is connected to an arm of an Aharonov-Bohm ring, dephasing can suppress backscattering, unlike in a ballistic channel where dephasing enhances backscattering. Additionally, dephasing reduces the amplitude of Aharonov-Bohm oscillations.

We also analyze the I-V characteristics of a one-dimensional channel with a dephasing center and a backscatterer (modeled by a weak hopping amplitude). We find that dephasing along with broken inversion results in rectification. This setup exhibits rectification, with the diode effect coefficient oscillating with bias.   The coefficient can switch from positive to negative depending on the distance between the dephasing center and the backscatterer. We provide explanations for the peaks and valleys in the diode effect coefficient as a function of bias.

While rectification is impossible in coherent two-terminal noninteracting metallic systems due to the unitarity of the scattering matrix, it can occur in superconducting systems with coherent scattering when time reversal and inversion are broken~\cite{soori2024bam}. Additionally, rectification is a known phenomenon in four-terminal noninteracting setups~\cite{song98}. We show that dephasing by B\"uttiker probe can cause diode effect in inversion broken metallic transport channel. 
Since the probe acts a measurement device, our results also provide a glimpse of how the quantum dynamics of these systems are affected under (local) continuous monitoring of charge density~\cite{Kohler2014,Marco2024}.  

\acknowledgments
 AS thanks  SERB Core Research grant (CRG/2022/004311) and University of Hyderabad Institute of Eminence PDF for funding. 
 UK was supported by fellowships from the Israel Science Foundation (ISF, Grant No.~993/19), the US-Israel Binational Science Foundation (BSF Grant No.~2016130), and the NSF-BSF Grant No.~2018726.

\bibliography{refdeco}

\end{document}